\documentclass[sigconf]{acmart}

\renewcommand\footnotetextcopyrightpermission[1]{} 
\usepackage[normalem]{ulem}
\usepackage{algorithm}
\usepackage{algorithmicx}
\usepackage{algpseudocode}
\usepackage{array}
\usepackage{xcolor}
\usepackage{todonotes}

\usepackage{hyperref}

\setcopyright{none}

\usepackage{listings}

\usepackage{soul,color}

\sethlcolor{white}

\title{CuLDA\_CGS: Solving Large-scale LDA Problems on GPUs}

\begin{document}

\author{Xiaolong Xie$^\dagger$, Yun Liang$^\dagger$, Xiuhong Li$^\dagger$, Wei Tan$^\ast$\vspace{-0.3cm}}\vspace{-0.3cm}
\vspace{-0.3cm}
\affiliation{%
  \institution{$^\dagger$Center for Energy-Efficient Computing and Applications, EECS, Peking University, Beijing, China }
  \city{$^\ast$Citadel LLC, Chicago, IL}
}
\email{{xiexl\_pku, ericlyun, lixiuhong}@pku.edu.cn, weitan@ieee.org}




\begin{abstract}

Latent Dirichlet Allocation(LDA) is a popular topic model. Given the fact that the input corpus of LDA algorithms consists of  millions to billions of tokens, the LDA training process is very time-consuming, which may prevent the usage of LDA in many scenarios, e.g., online service. GPUs have benefited modern machine learning algorithms and big data analysis as they can provide high 
memory bandwidth and computation power. Therefore, many frameworks, e.g. TensorFlow, Caffe, CNTK, support to use GPUs for accelerating the popular machine learning data-intensive algorithms. However, we observe that LDA solutions on GPUs are not satisfying. 

In this paper, we present CuLDA\_CGS, a GPU-based efficient and scalable approach to accelerate large-scale LDA problems. CuLDA\_CGS is designed to efficiently solve LDA problems at high throughput. To it, we first delicately design workload partition and synchronization mechanism to exploit the benefits of multiple GPUs. Then, we offload the LDA sampling process to each individual GPU by optimizing from the sampling algorithm, parallelization, and data compression perspectives. Evaluations show that compared with state-of-the-art LDA solutions, CuLDA\_CGS outperforms them by a large margin (up to 7.3X) on a single GPU. CuLDA\_CGS is able to achieve extra 3.0X speedup on 4 GPUs. The source code is publicly available on \url{https://github.com/cuMF/CuLDA_CGS}.

\end{abstract}





\maketitle


\section{Introduction}

Machine learning algorithms are now widely used for automatic big data analysis~\cite{kdd16node}. Latent Dirichlet Allocation (LDA)~\cite{lda2003,sigir99plsi} is a popular Bayesian topic model and it's been demonstrated to be practical in real applications. The research community has conducted a variety of studies to optimize LDA and broaden it's usage scope~\cite{scc17lda,kdd16streamlda,kdd12context,kdd16topicmodel,kdd16shorttext,kdd17topic}. In general, existing studies could be categorized into two streams. One stream is \textit{algorithmic optimization} that aims to improve the accuracy or reduce the number of training iterations to converge~\cite{kdd13cgs}. In this paper, we focus on another stream, \textit{system optimization}~\cite{yahoo12,canny2013bidmach,vldb16warp,kdd14alias,asplos17saber,kdd15bayeslda,kdd15pet,kdd16dislda,kdd09sparse,www15nomad,www15light,vldb17yu,kdd13bigdata}. Our goal is to achieve high system throughput to minimize the cost of large-scale LDA training.

A typical LDA problem needs to infer $K$ topics from a corpus that is consisted  of up to millions of documents and up to billions of tokens. To achieve the goal, various training algorithms have been proposed~\cite{kdd09sparse,kdd13cgs}. At each step of the training, one token is selected from the corpus and a new topic is assigned to the token. In general, it requires hundreds of iterations to converge to an accurate model. The sampling complexity is $O(K)$ and given that the input data size is large, similar to other big data analysis applications~\cite{kdd13graph,kdd14batch,kdd16cloud,ppopp15gun}, the LDA training process is very time-consuming. The total elapsed time ranges from hundreds of seconds to tens of hours~\cite{kdd17topic}.

Various CPU-based LDA optimization techniques, e.g. sparsity-aware sampling~\cite{kdd09sparse}, computing complexity optimization~\cite{kdd14alias,www15light}, memory optimization~\cite{kdd15bayeslda,vldb16warp,yahoo12}, distributed computing~\cite{www15nomad,vldb17yu,kdd16dislda,kdd15pet}, have been proposed. However, the performance of existing LDA solutions are still not satisfying. The LDA training is mainly limited by the memory bandwidth. For example, the typical LDA solution only performs 0.27 arithmetic operation for each byte data loaded from the memory\footnote{Detailed analysis is shown in Section~\ref{sec:mot}.}. Single-node CPU solutions mainly rely on caches to improve the memory bandwidth~\cite{vldb16warp}. However, the increasing data size makes the cache performance sub-optimal~\cite{hpdc17mf}. Distributed LDA solutions does not face the cache problems, however, the limited network speed becomes the performance bottleneck as the LDA algorithm requires frequent model data communication. Other big data applications face the similar problems, too. Therefore, systems that are equipped with hardware accelerators ( e.g., GPU, TPU~\cite{isca17tpu}, ASIC~\cite{asplos14diannao}) are employed for these big data applications. In this paper, we use GPUs. GPUs are many-core architectures and it provides much higher memory bandwith compared to CPUs.

However, due to the distinct architecture design, simply porting existing CPU-based or distributed system-based LDA solutions to GPUs can not deliver good performance~\cite{hpdc17mf}. GPU-based LDA solutions has also been studied by other researchers~\cite{canny2013bidmach,asplos17saber}. However, we observe that existing solutions are not able to fully release the horse power of GPUs nor scalable to multiple GPUs. Several challenges remains. First, the LDA algorithms are basically irregular. It is very difficult to achieve high resource utilization on GPUs~\cite{hpdc17mf}. Second, the inter-GPU synchronization mechanism has high impact on the system throughput. A scalable multi-GPU solution for large-scale LDA problems is essential. To solve the above challenges, we present \textit{CuLDA\_CGS}, an efficient and scalable LDA solution on GPUs. Our design goal is to solve large-scale LDA problems with one single machine and achieve comparable or even better performance than distributed systems. To this, we not only focus on the single-GPU performance, but also the multi-GPU (single machine) scalability.


We contribute to the state-of-the-art machine learning and big data analysis research community as following,
\begin{itemize}
\item We conduct comprehensive analysis on LDA training algorithms and identify that LDA training is memory bound.
\item We propose to employ systems that consist of GPUs for accelerating LDA training and develop CuLDA\_CGS, an efficient LDA training algorithm to fully utilize the GPU hardware resources. CuLDA\_CGS is well-optimized in many aspects, including but to limited to memory resources utilization, workload partition, and multi-GPU scalability.
\item We evaluate CuLDA\_CGS on different GPU architectures using real data sets. Evaluations demonstrate that CuLDA\_CGS is able to achieve high throughput and scalable to different GPU architectures and different number of GPUs. We also compare it with the state-of-the-art LDA solutions and CuLDA\_CGS outperforms existing solutions by a large margin (up to 7.3X).
\item CuLDA\_CGS achieves an extra 1.9X, 3.0X speedup on 2 GPUs and 4 GPUs, respectively. To the best of our knowledge, CuLDA\_CGS is the first LDA solution that is able to scale to multiple GPUs.
\end{itemize}

The remainder of this paper is organized as follows. We describe the LDA and GPU background knowledge in Section~\ref{sec:back}. In Section~\ref{sec:mot}, we show the bottleneck analysis of LDA algorithm and motivate our study. In section~\ref{sec:overview}, we show the system overview of CuLDA\_CGS. Section~\ref{sec:data} describes the details of synchronization mechanism that exploits the computation horsepower of multiple GPUs, Section~\ref{sec:model} describes how does CuLDA\_CGS effectively executes LDA samplings and model update on GPUs. Finally, Section~\ref{sec:con} concludes this paper.

\section{Background}\label{sec:back}

\subsection{LDA Model}

\begin{figure}[h]
\centering
\includegraphics[scale=0.6,angle=0]{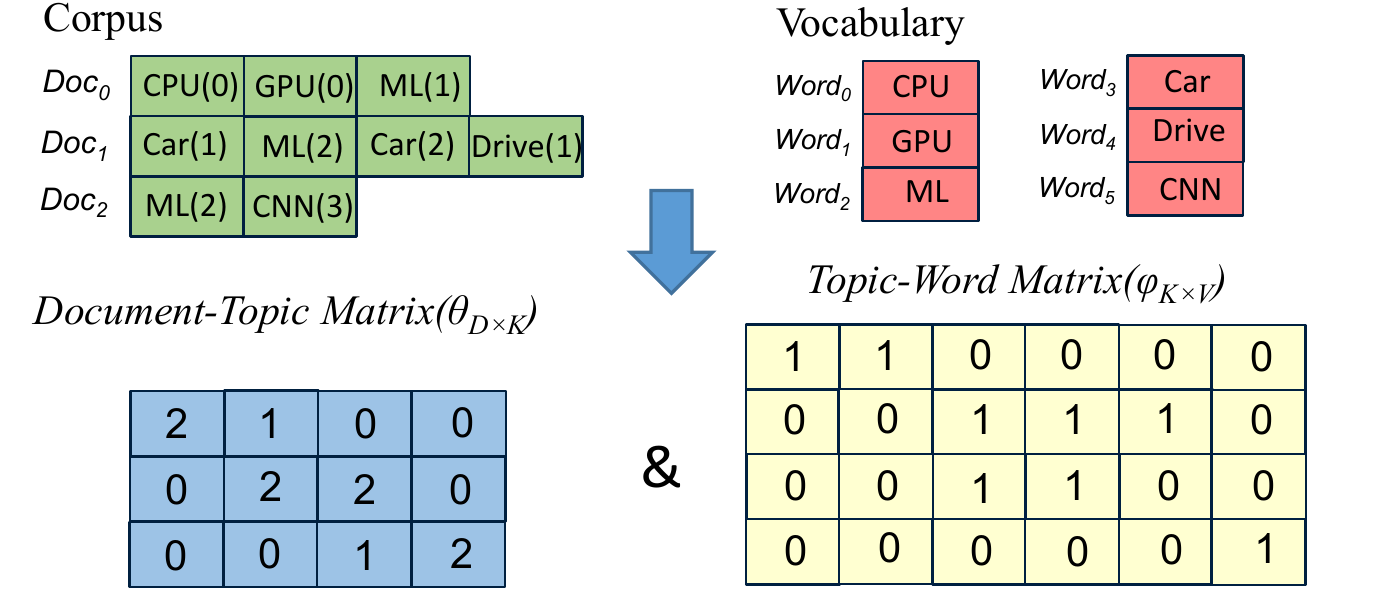}\vspace{-0.3cm}
\caption{A pseudo example of LDA application($K=3$, $D=3$, $V=6$).}\label{figure:doc}
\end{figure}

\textit{Topic Model} is a statistical model that is used to infer $K$ topics from a corpus. LDA model~\cite{lda2003} is an efficient solution for topic model. It is widely used in text analysis and other popular applications~\cite{scc17lda,kdd16shorttext}. The input corpus is a collection of documents which is composed of $D$ documents($Doc_0, Doc_1, ..., Doc_{D-1}$). Each document is a group of tokens. A token is an instance of a word and the vocabulary size is $V$($Word_0, Word_1,...,Word_{V-1}$). Note that one word may appear multiple times in one document. Initially, each token is randomly assigned with a topic. For a given document set ($d\in[0,D)$) and vocabulary($v\in[0,V)$ ), the process of LDA training is to infer a handful of topics ($k\in [0,K)$ ) and to infer the topic distribution of each document. Figure~\ref{figure:doc} shows an pseudo example of LDA application. More details about LDA model are provided in ~\cite{lda2003}.

The output of LDA model is:
\begin{itemize}
\item Document-topic matrix $\theta_{D\times K}$, where $\theta_{d,k}$ represents the number of words of topic $k$ in document $d$.
\item Topic-word matrix $\phi_{K\times V}$, where $\phi_{k,v}$ represents the number of occurrences of word $v$ of topic $k$ in all documents.
\end{itemize}

There exists several solutions to train matrices (model) $\theta$ and $\phi$. In this paper, we adopt the \textit{Collapsed Gibbs Sampling} (CGS) algorithm~\cite{kdd09sparse,kdd13cgs}. CGS iteratively picks up one token from the corpus and reassigns a new topic for this token following the multinomial possibility distribution:
\begin{equation}\label{eq:dis}
p(k) = \frac{(\theta_{d,k} + \alpha)(\phi_{k,v} + \beta) }{\sum_{v\in[0,V)} \phi_{k,v} + \beta V}
\end{equation}
where p(k) represents the possibility of choosing topic $k$. $\alpha$ and $\beta$ are hyper parameters. In this paper, same with the previous paper~\cite{vldb16warp}, we set $\alpha$ as $K/50$ and $\beta$ as 0.01. We refer a full pass of all tokens as a iteration. We term this step as \textit{sampling}. After each iteration, the model $\theta_{D\times K}$ and $\phi_{K \times V}$ are updated and a new iteration is started until the model converges. 


The training typically requires hundreds of iteration to converge. Given that the input corpus is often large and K ranges from 1k to 10k, the training time is very long. To quantify the performance, we use the number of processed tokens per second ($\#Tokens/sec$) as the performance metric:

\begin{equation}\label{eq:metric}
\#Tokens/sec = \frac{\#Tokens \times \#Iterations}{Elapsed Time} 
\end{equation}

\subsection{GPU Architecture and Programming Interface}

\begin{figure}[h]
\centering
\includegraphics[scale=0.45,angle=0]{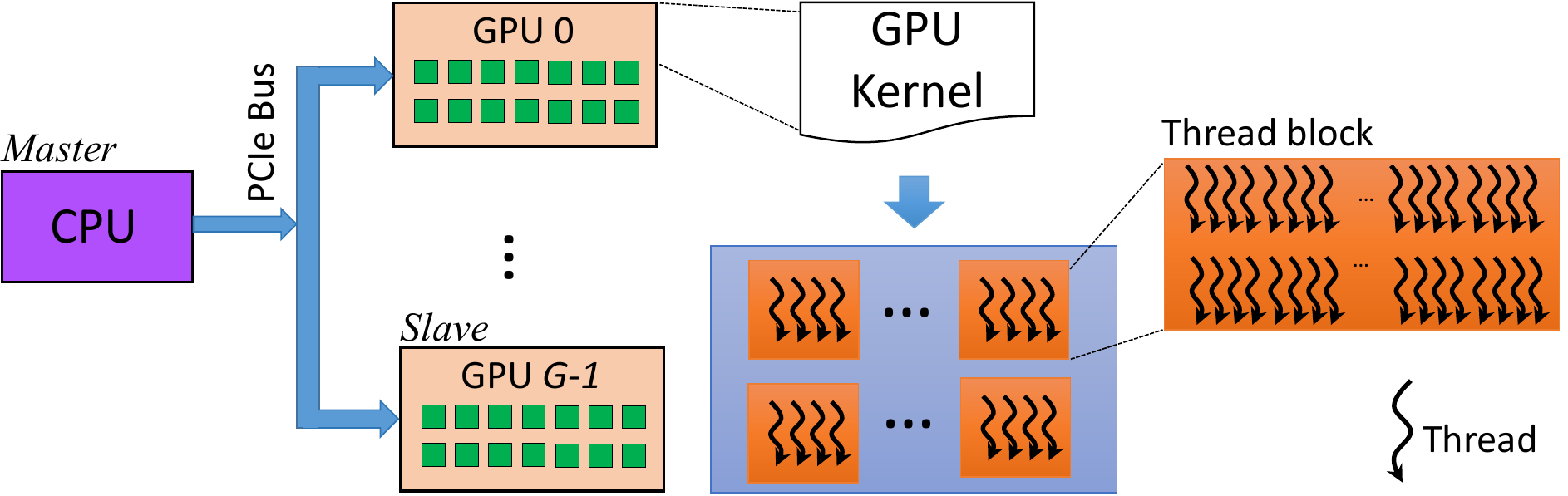}\vspace{-0.3cm}
\caption{GPU architecture and programming model.}\label{figure:arch}
\end{figure}
We use NVIDIA CUDA programming interface~\cite{guide}. Figure~\ref{figure:arch} shows the architecture and programming model of GPUs. The heterogeneous system is composed of CPUs and multiple GPUs. The CPUs and GPUs are connected via PCIe bus and work in a master-slave mode. The application explicitly manages the data transfer between processors and sends control commands to GPUs. Using CUDA terminology, a GPU function is termed as a \textit{kernel}. By default, a GPU executes one kernel at a time. When a kernel is launched to GPU, it spawns hundreds of thousands of threads. Those threads share the same instruction stream. Each thread is assigned a unique ID, therefore, they operate on different input data elements. 

The threads are organized into thread blocks. Threads within one thread block can synchronize through intrinsic API and more importantly, they can exchange data using the \textit{shared memory}(aka soft-ware managed cache). Shared memory can provide much shorter latency and higher throughput than memory as it's on-chip. One GPU can execute hundreds of thread blocks concurrently. Thread blocks with smaller IDs are issued first. At the hardware-level, consecutive 32 thread is grouped into one warp (64 on AMD GPUs). Threads in one warp are executed as a vector group and they can exchange data through register file(faster than shared memory).

\section{Motivation}\label{sec:mot}

In this Section, we present our research motivation. Fully utilizing the tremendous hardware resources on GPUs is not a trivial task. In Section~\ref{sec:char}, we give the characterization of LDA sampling algorithm and the benefits of using GPUs. In Section~\ref{sec:challenge}, we present the design challenges.

\subsection{Characterization}\label{sec:char}

To find the performance bottleneck of LDA, we analyze the resource consumption of each LDA sampling. Note that our characterization is independent of platform. We use the roofline model to simplify our analysis~\cite{acm09roofline}. Roofline model is often used to identify the performance bottleneck on parallel processors. It relys on a metric, $Flops/Byte$

\begin{equation}\label{eq:com2mem}
{ Flops/Byte = \frac{\#FloatingPointOps}{\#MemoryOps(Byte)} }
\end{equation}

to represent the computation to memory ratio of an application. If the $Flops/Byte$ is larger than the peak FLOPS to peak memory bandwidth ratio of the processor, the application is bound by computation and otherwise, it's bound by memory. We examine the LDA algorithm (Algorithm~\ref{algo:lda}) and calculate the $Flops/Byte$ for each step of LDA sampling.  We use 32-bit integer and 32-bit floating point data representation.  Document-topic matrix  $\theta$ is stored in CSR format\footnote{\url{https://en.wikipedia.org/wiki/Sparse_matrix}}. Table~\ref{table:byte} shows the results. On average, the {Flops/Byte} of LDA is 0.27. We use the latest Volta architecture from NVIDIA as our experiment platforms (Table~\ref{table:platform}). We use the CPU used in the Volta platform as comparison, it provides up to 470 GFLOPS and 51.2 GB/s theoretical peak performance(470/51.2=9.2). We  conclude that LDA is bound by memory bandwidth. 

\begin{table}[h]
\caption{Flops/Byte of each steps of one LDA sampling. $K_d$ is the number of non-zero elements of $\theta_{d,*}$.}\label{table:byte}\vspace{-0.3cm}
\small
\centering
\begin{tabular}{|c|c|c|}
\hline
  \textbf{Step}   & \textbf{Formulas}             & \textbf{Values} \\ \hline
  Compute S       & $4*K_d/(3*Int*K_d)$           & 0.33 \\ \hline
  Compute Q       & $2*K/(2*Int*K)$               & 0.25 \\ \hline
  Sampling from $p_1(k)$ & $6*K_d/((3*Int+2*Float)*K_d)$ & 0.30  \\ \hline
  Sampling from $p_2(k)$& $3*K/((2*Int+2*Float)*K)$     & 0.19 \\ \hline
\end{tabular}
\end{table}

\subsection{Challenges}\label{sec:challenge}

Modern CPUs rely on caches to improve the effective memory bandwidth of applications. Previous LDA solutions~\cite{vldb16warp} has focused on optimizing the cache performance. However, given that the input of LDA applications is often large and keeps increasing, the working set size can not fit into CPU's caches. Therefore, CPU-based LDA solutions is not scalable for big data. One solution is to distribute the data set on distributed systems to improve the cache performance~\cite{www15nomad,vldb17yu,kdd16dislda,kdd15pet}. However, as the processors on different nodes need to synchronize the model at each iteration, the performance is limited by the network bandwidth. Due to the above limitations, in this paper, we propose to use GPUs to solve large-scale LDA problems. 

\begin{figure*}[ht]
\centering
\includegraphics[scale=1,angle=0]{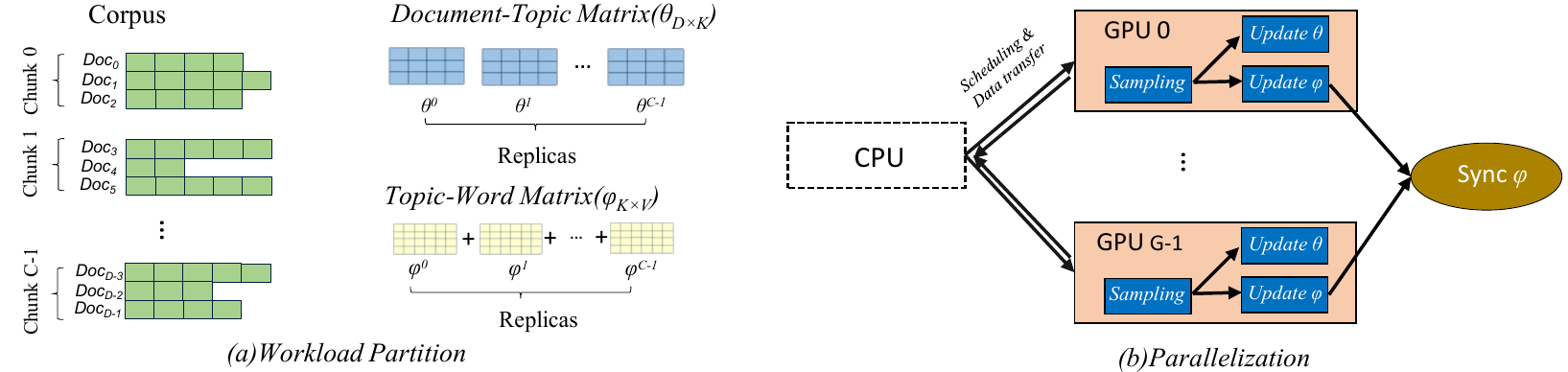}\vspace{-0.3cm}
\caption{The design overview of CuLDA\_CGS. The input corpus is partitioned into $C$ chunks.}\label{figure:overview}
\end{figure*}

We show the overview of the targeted system in Figure~\ref{figure:overview}. The CPUs are responsible for data preprocessing and workload management. After the data preprocessing, the CPUs are responsible for distributing the workloads to GPUs and collecting the results when the GPUs finish the execution. There are two benefits of using such a system. On one hand, each GPU processor is able to provide more resources than one CPU processor. For example, one NVIDIA Volta V100 GPU can provides up to 900GB/s peak memory bandwidth and 1,400 GFLOPS peak single-precision speed. Moreover, GPUs provide on-chip memory resources(e.g., shared memory) to accelerate memory accesses. On other hand, The processors are connected via PCIe 3.0 bus that provides up to 16GB/s bandwidth. The most-recent NVLink can provides up to 300GB/s connection bandwidth~\cite{dgx}. In comparison, the bandwidth of used interconnection in~\cite{vldb17yu} is only 10 Gb/s .
In conclusion, the system shown in Figure~\ref{figure:arch} is suitable for communication-rich applications like LDA.


Though the benefits of using GPU-based system is clear, there are still many challenges when designing an efficient GPU solution. First, to achieve the optimal performance, we need to optimize the single-GPU sampling speed. GPUs are many-core architectures, it's necessary to launch tens of thousands of concurrent threads to saturate one GPU processor. In comparison, dozens of threads is enough to saturate one CPU processor. Therefore, we need to design extreme light-weight workload scheduling and sampling algorithms for GPUs. At the meantime, we also need to carefully utilize the on-chip memory resources on GPUs to improve the effective memory bandwidth. Second, at each iteration, after sampling, we need to update the model. The original update algorithm is irregular and not suitable for GPUs~\cite{irregular}. We have to transform the algorithm to make sure that the model update step is not the performance bottleneck. Third, scaling to large-scale data set and multiple GPUs is not a trivial task. The memory capacity of GPUs is smaller than CPUs, simply allocating space for all of the data in GPU's memory is infeasible for large-scale LDA problems. At the meantime, as the GPUs is more powerful than CPUs, according to Amadhl's law, the synchronization overhead becomes critical for overall performance. If the synchronization between CPUs and GPUs is not well-optimized, we are not able to benefit from multiple GPUs. We need to design an efficient workload partition and communication mechanism.

\section{Overview}\label{sec:overview}

To overcome the challenges, we design CuLDA\_CGS and present its overview in this section. Figure~\ref{figure:overview}(a) shows the workload partition mechanism of CuLDA\_CGS. We partition the input corpus into multiple chunks. Correspondingly, the document-topic matrix $\theta_{D\times K}$ and topic-word matrix $\phi_{K\times V}$ are also partitioned into multiple replicas. We will discuss the decision of number of chunks in Section~\ref{sec:data}. 

Basically, there are two workload partition policy, partition-by-document and partition-by-word. For the partition-by-word policy, $\theta_{D\times K}$ is spatially partitioned and therefore, $\theta_{D\times K}$ is spatially partitioned while $\phi_{V\times K}$ equals to the summation of all replicas. Therefore, after the sampling, we only need to synchronize each replica of $\phi_{K\times V}$. For the partition-by-word policy, we only need to synchronize the replicas of $\theta_{D\times K}$. Consider $D$ is often several orders of magnitude greater than $V$, synchronize $\theta_{D\times K}$ is more expensive than $\phi_{V\times K}$. Therefore, we select the partition-by-document policy. Different documents have different number of tokens. To avoid load imbalance, the corpus is evenly partitioned by number of tokens, instead of number of documents.

Figure~\ref{figure:overview}(b) shows the parallelization schema. Originally, all of the corpus chunks and the replicas of $\theta_{D\times K}$ and $\phi_{K \times V}$ resides in the main memory of CPU. The CPU is responsible for the workload scheduling and data transfers for GPUs. When a data chunk is scheduled to one GPU, the GPU samples the corpus chunk, update $\theta$ replica, and update $\phi$ replica. After that the GPUs synchronize matrix $\phi$ and the CPU schedules new workloads for the GPU. We describe the details of workload scheduling and $\phi$ synchronization algorithms in Section~\ref{sec:data}. We describe how each GPU efficiently samples the tokens and the details of model update algorithms in Section~\ref{sec:model}.

\section{Parallelization Scheme}\label{sec:data}

We describe how to design the workload scheduling algorithm and model synchronization algorithms for CuLDA\_CGS in this Section.

\algnewcommand\algorithmicinput{\textbf{Input:}}
\algnewcommand\Input{\item[\algorithmicinput]}
\algnewcommand\algorithmicoutput{\textbf{Output:}}
\algnewcommand\Output{\item[\algorithmicoutput]}
\algnewcommand{\LineComment}[1]{\State \(\triangleright\) #1}
\begin{algorithm}[h]
\small
\caption{Workload Scheduling Algorithm.}\label{algo:schedule}
\begin{algorithmic}[1]
\Input Corpus chunks, $\theta_{D\times K}$, $\phi_{K\times V}$, $C$=$M\times G$,\#Ite
\If{M==1}
\State Call WorkSchedule1();
\Else
\State Call WorkSchedule2();
\EndIf
\Procedure{WorkSchedule1()}{}
    \For{$i \in [0,C)$}
        \State DataTransfer($Chunk^i$,$CPU$ $\rightarrow$ $GPU^i$)
    \EndFor
    \For{$ite \in [0,\#Ite)$}
        \For{$i \in[0,C]$} in parallel
        \State $GPU^i$.run()
        \EndFor
        \State Reduce($\phi^0,$...,$\phi^C-1$)
    \State Broadcast($\phi^0$)
    \EndFor
    \State DataTransfer($\phi^0$, $GPU^0$ $\rightarrow$ $CPU$)
    \For{$i \in [0,C)$}
        \State DataTransfer($\theta^i$,$GPU^i$ $\rightarrow$ $CPU$)
    \EndFor
\EndProcedure

\Procedure{WorkSchedule2()}{}
    \For{$ite \in [0,\#Ite)$}
        \For{$i \in[0,C]$} in parallel
            \For {$m \in[0,M]$}
                \State DataTransfer($Chunk^{M\times i + m}$, $CPU$ $\rightarrow$ $GPU^i$)
                \State DataTransfer($\theta^{M\times i + m}$, $CPU$ $\rightarrow$ $GPU^i$)
                \State $GPU^i$.run()
                \State DataTransfer($\theta^{M\times i + m}$, $GPU^i$ $\rightarrow$ $CPU$)
            \EndFor
        \EndFor
        \State Reduce($\phi^0,$...,$\phi^C-1$)
        \State Broadcast($\phi^0$)
    \EndFor
    \State DataTransfer($\phi^0$, $GPU^0$ $\rightarrow$ $CPU$)
\EndProcedure
\end{algorithmic}
\end{algorithm}

\subsection{Scheduling Algorithm}\label{sec:multi}
 
As shown in Figure~\ref{figure:overview}, to exploit the resources on $G$ GPUs and avoid load imbalance, we evenly partition the corpus into $C$ chunks and $C=M \times G$($M$ $\ge$ 1). At each iteration, the chunks are scheduled to GPUs in a round-robin order. Chunk $i$ is scheduled to GPU $i\%G$ and the chunks with smaller IDs are scheduled first. After all GPUs finish there execution, the model update kernels are called. The scheduling algorithm continues until the number of given iterations is reached.


The ideal choice is $M$ = 1. If $M$ = 1, the data transfer between CPU and GPUs only happens at the start and the end of the computation. Given that a LDA algorithm needs hundreds of iterations($\#Ite$) to converge, the amortized data transfer overhead is negligible. However, as discussed before, one main drawback of using GPUs is the device memory capacity. A typical GPU has only 12GB-16GB memory, which is much smaller than CPU's memory capacity. Thus, when deciding the value of $M$, we need to make sure that the one GPU's memory can accommodate at least one data chunk. 

We show the detailed workload scheduling algorithm in Algorithm~\ref{algo:schedule}. When $M$ = $1$, CuLDA\_CGS calls scheduling procedure WorkSchedule1()(Line 1-2). The declaration of procedure WorkSchedule1() is shown at Line 6-21. At the start of the LDA training, CuLDA\_CGS transfers to data to GPUs(Line 7-9) and after the training (Line 10-13), the CPU collectes the trained model from all GPUs (Line 17-20). The training needs $\#Ite$ iterations to finish. At each iteration, CuLDA\_CGS calls each GPU in parallel(Line 11-13) and after all GPUs finish their execution, the CPU synchronizes the topic-word matrix $\phi$. The details of the synchronization algorithm are shown in Section~\ref{sec:sync}.

When $M$ > 1, CuLDA\_CGS calls scheduling procedure WorkSchedule2()(Line 3-4) and the declaration of procedure WorkSchedule2() is shown at Line 22-36. The main different between WorkSchedule2() and WorkSchedule1() is, the data transfer between CPU and GPUs happens at each iteration(Line 26-29). Despite that the PCIe bandwidth is higher than Ethernet network, the transfer overhead is still a bottleneck for the performance. To minimize the data transfer overhead, on one hand, we compress the data sizes of data chunks and models, on other hand, we overlap the data transfer and the computation. We will discuss how to compress the data size in Section~\ref{sec:model}. As for the overlap, the main idea is to pipeline the loop(Line 25-30) and overlap the transfer of $(m+1)^{th}$ loop with the computation of $m^{th}$ loop. We employ the GPU's \textit{stream} interface. To overlap the computation and memory transfer, we need to allocate two data chunks in the GPU's memory. Therefore, when choosing value for $M$, we need to make sure the GPU's memory is able to accommodate two data chunks.

\subsection{Model Synchronization Algorithm}\label{sec:sync}

As discussed in Section~\ref{sec:multi}, at each iteration, after each GPU finishes it's computation, we need to synchronize the topic-word matrix $\phi$ as follows,
\begin{equation}
\phi = \phi^0 + ... \phi^{G-1}
\end{equation}

\begin{figure}[h]
\centering
\includegraphics[scale=0.28,angle=0]{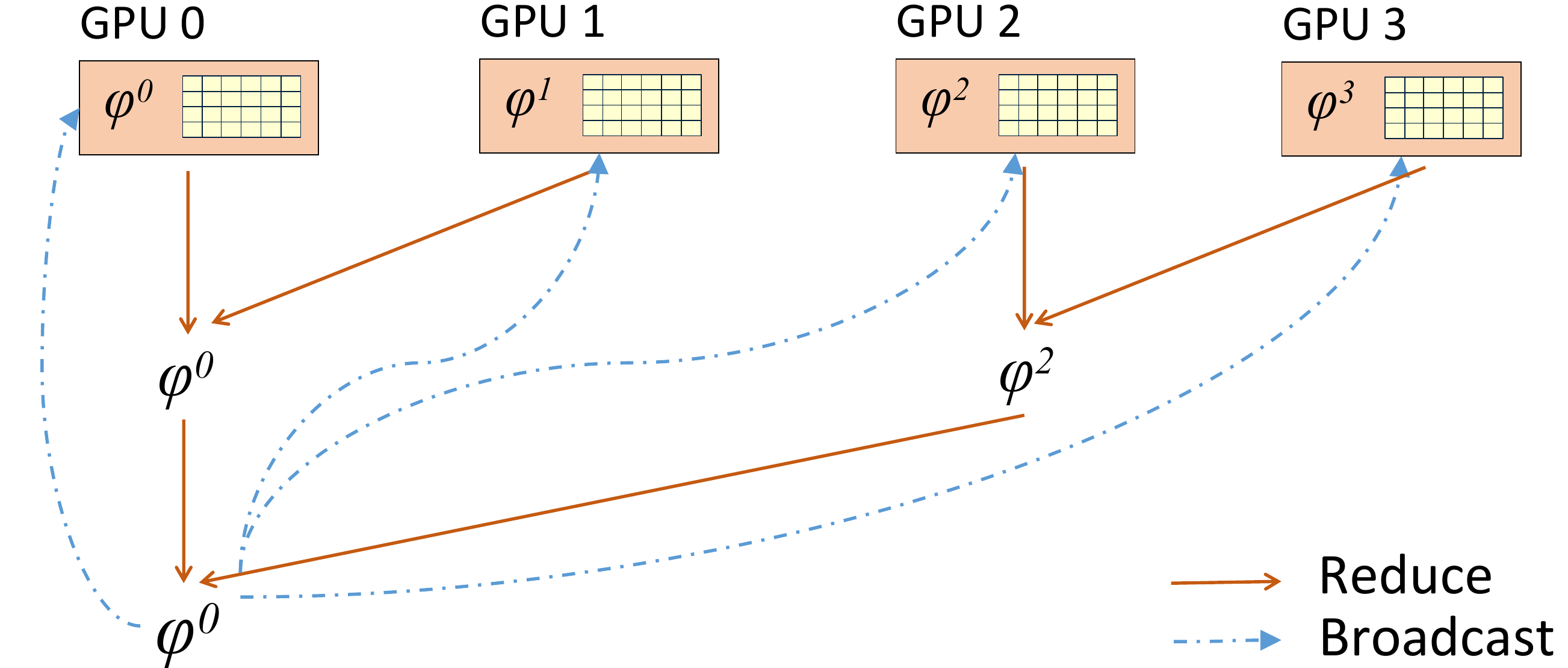}\vspace{-0.3cm}
\caption{The model synchronization algorithm for $\phi_{K\times V}$ is composed of two operations, \textit{reduce} and \textit{broadcast}.}\label{figure:reduce}
\end{figure}

An intuitive solution is to transfer all $\phi$ replicas to CPU and add all the replicas. However, we observe it's not optimal as the CPU is slower than GPUs in terms of matrix adding. Therefore, in this step, we only use GPUs to perform the synchronization. To minimize the overhead, we split the synchronization into two operations, \textit{reduce} and \textit{broadcast}. We shows how CuLDA\_CGS perform the synchronization in Figure~\ref{figure:reduce}. We use four GPUs in the psuedo case. In this case, it requires two iterations to finish the reduce operation. In the first iteration, GPU 1 transfers $\phi^1$ to GPU 0 and GPU 3 transfers $\phi^3$ to GPU 2. After the data transfers, GPU 0 sums $\phi^0$ and $\phi^1$ up and GPU 2 sums $\phi^2$ and $\phi^3$ up, respectively. At the second step, GPU 2 transfers $\phi^2$ to GPU 0 and GPU 0 sums $\phi^0$ and $\phi^2$ up. $\phi^0$ is the result. After the reduction, we broadcast the result($\phi^0$) to all GPUs. Note that reductions in the same iteration is performed in parallel, therefore, the computation complexity of reduction is $logG$. By doing so, our algorithm is still efficient when scaling to multiple GPUs.


\section{GPU Kernel Optimization}\label{sec:model}
For each GPU, its computation consists of three kernels, sampling, update $\theta$, and update $\phi$ as shown in Figure~\ref{figure:overview}. Here, we discuss how to optimize the sampling kernel in Section~\ref{sec:sample} and how to optimize the update kernels in Section~\ref{sec:update}.

\subsection{Sampling Optimization}\label{sec:sample}
In the following, we optimize one LDA sampler (Section~\ref{sec:sampler}) , parallelize thousands of samplers (Section~\ref{sec:para}), and optimize the memory accesses (Section~\ref{sec:compress}).

\subsubsection{Sampler Design}\label{sec:sampler}

Our sampler is based on the \textit{sparsity-aware sampling}~\cite{kdd09sparse}. As shown in Eq.~\ref{eq:dis}, at each step, the LDA sampler computes the vector $p(k)$ at length $K$ and generates a new topic. The computing complexity is $O(K)$. To reduce the complexity, sparsity-aware LDA sampling~\cite{kdd09sparse} is widely adopted. The basis observation of sparsity-aware sampling is that the doc-topic matrix $\theta_{D\times K}$ is often sparse because

\begin{equation}
DocLen_{d} = \sum_{k\in[0,K)}\theta_{d,k}
\end{equation}

where $DocLen_d$ is the number of tokens in the $d_{th}$ document. Given the fact that $DocLen_{d}$ is often much smaller than $K$, the number of non-zero elements in $d_{th}$ line of $\theta$ is often much smaller than $K$. Hence, $\theta_{D\times K}$ is a sparse matrix. To exploit this feature, the $p(k)$(Eq.~\ref{eq:dis}) is decomposed to:
\begin{equation}\label{eq:pk}
\begin{split}
p(k) &= p_1(k) + p_2(k) \\
p_1(k) &= \frac{\theta_{d,k}(\phi_{k,v} + \beta)}{\sum_{v\in[0,V)} \phi_{k,v} + \beta V} \\
p_2(k) &= \frac{\alpha (\phi_{k,v} + \beta)}{\sum_{v\in[0,V)} \phi_{k,v} + \beta V}
\end{split}
\end{equation}

We find out that $p_1(k)$ is a sparse vector while $p_2(k)$ is dense. To exploit this feature, before the sampling, we first compute $S$ and $Q$,

\begin{equation}
S = \sum_{k\in [0,K)}^{}p_1(k), \ \ \ \ Q = \sum_{k\in [0,K)}^{}p_2(k)
\end{equation}

As $S$ is the sum of $p1(k)$, which is sparse, we can exploit the sparsity to minimize the computation overhead of S as we only need to consider p1(k) where $\theta_{d,k}$ is non-zero. Then we generate a random value, $u\sim U(0,1)$. If $u<$ $S/(S+Q)$, then we sample a new topic $p_z$  $\propto multi(p_1(k))$. Otherwise, we sample a new topic $p_z$ $\propto multi(p_2(k))$. As $\theta_{D\times K}$ is sparse, the computation complexity is reduced.



\begin{figure}[h]
\centering
\includegraphics[scale=0.9,angle=0]{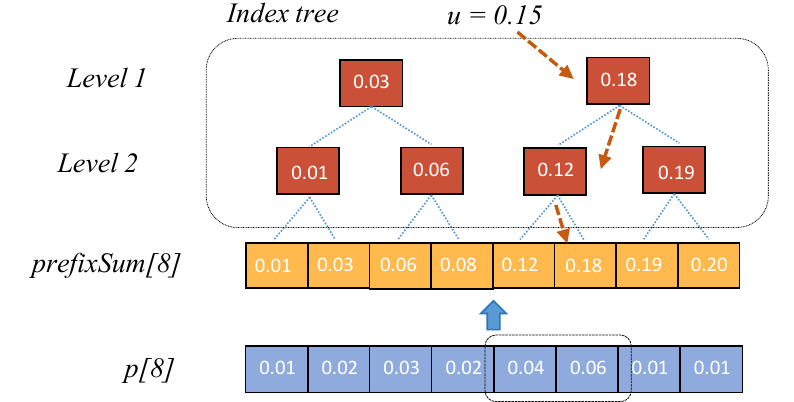}\vspace{-0.3cm}
\caption{Idea of tree-based sampling.}\label{figure:tree}
\end{figure}

GPUs provides shared memory as a software-managed cache. If well-managed, the shared memory is able to provide high memory throughput. In the sparsity-aware sampling, the possibility array $p_1(k)$ and $p_2(k)$ are accessed twice, computing $S \& Q$ and sampling. However, the shared memory is not large enough to accommodate the entire array. We exploit \textit{tree-based} structure to reuse the data. Figure~\ref{figure:tree} illustrates the idea. In the pseudo case, we need to sample a topic from $p[8]$. To it, we first compute the prefix summation of $p[8]$($prefixSum[8]$). Then we randomly generate u $\propto U(0,p[7])$. To get the topic, we need to find the minimal $k$ that makes $p[k]>u$. By doing so, we transform the sampling to a search problem. We construct a binary search tree to find the result. The benefit of tree-based sampling is two-fold. The first is, the sampling overhead is reduced. The second is, the index tree is small enough to fit into shared memory. At the sampling step, we only need to access the data in the circles. More importantly, the index tree is in the fast shared memory, only the two elements of $p[8]$ are in the memory. Therefore, the memory footprint is minimized.

We put the sparsity-aware sampling and tree-based sampling together. Algorithm~\ref{algo:lda} shows the design of one sampler. In the shown algorithm, we only present one thread. The length of the operations in Algorithm~\ref{algo:lda} is $K$ or $K_d$. GPUs are SIMD(single-instruction, multiple data)processors, On NVIDIA GPUs, consecutive 32 GPU threads are grouped into one SIMD group, aka warp. Threads in one warp can works together as a vector machine. CuLDA\_CGS uses one warp to process one LDA sampling at a time. We refer a warp as a sampler. Similarly, we use 32-way tree in the tree-based sampling .

\begin{algorithm}[H]
\small
    \caption{The design of LDA sampler.}\label{algo:lda}
    \begin{algorithmic}[1]
    \Procedure{LDAsampler()}{}
        \State get\_a\_token();
        \State compute(S) \& construct tree for $p_1(k)$
        \State compute(Q) \& construct tree for $p_2(k)$
        \State sample $u\backsim U(0,1)$
        \If{$u \le S/(S+Q)$} 
            \State {sample $p_z \propto multinomial(p_1(k)))$} 
        \Else
            \State {sample $p_z \propto multinomial(p_2(k))$} 
        \EndIf 
    \EndProcedure
\end{algorithmic}
\end{algorithm}

\begin{figure}[ht]
\centering
\includegraphics[scale=0.59,angle=0]{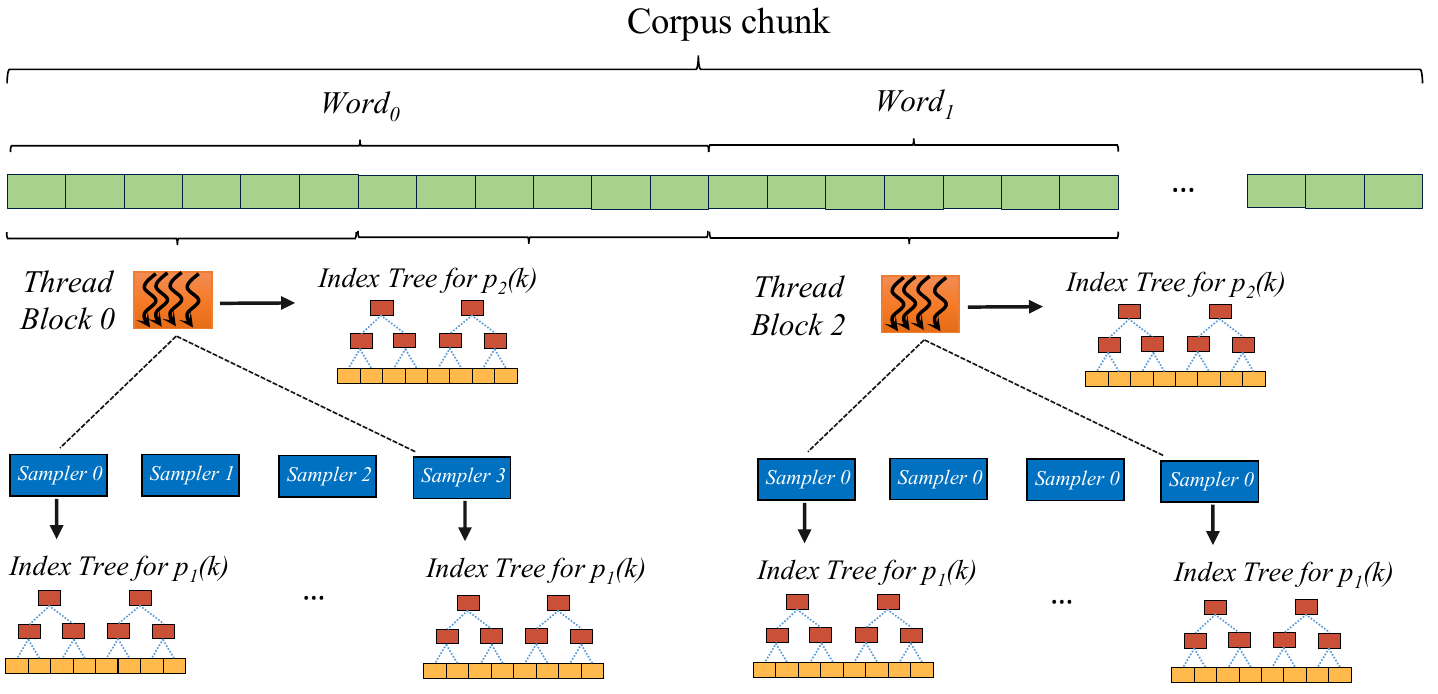}
\caption{The overview of LDA sampler parallelization algorithm used in CuLDA\_CGS.}\label{figure:sample}
\end{figure}

\subsubsection{Parallelization}\label{sec:para}
A GPU can execute tens of thousands of threads concurrently. Hence, running one sampler can not fully utilize the GPU. An intuitive parallelization mechanism is to launch one sampler for one token and let the GPU to determine the execution order. However, we observe that it's not optimal.

As addressed before, LDA is bound by the memory bandwidth. To maximize the memory bandwidth, we try to utilize the software-managed cache(aka \textit{shared memory}) on the GPUs~\cite{micro15crat}. The shared memory is managed by software and the cache hit rate for the allocated data is 100\%. Therefore, it's very important for memory bound applications. From Eq.~\ref{eq:pk}, we can find that tokens of the same word shares the same values of $p_2(k)$. Therefore, for the given corpus chunk, we sort the tokens in a word-first order. As the samplers in the same thread block could share data through shared memory, they can share the same index tree for $p_2(k)$. We let the samplers in the same thread block sample the tokens from the same word. We set the number of samplers in each thread block as 32, which is the allowed maximal value.  

We show the overview of our sampling parallelism algorithm in Figure~\ref{figure:sample}. The tokens assigned to one thread block are evenly partitioned by all samplers in the thread block. Thus, samplers in one thread block share the same index tree for $p_2(k)$. Words that have a lot of tokens are assigned to multiple thread blocks to avoid load imbalance. At the mean time, those words are assigned to thread blocks that have the smallest IDs to avoid long-tail effect. We let the GPU scheduler to determine the execution order of thread blocks and by default, the GPU scheduler schedules those block first. We also apply the tree-based sampling for the sampling from $p_1(k)$. Line 4 and Line 9 in Algorithm~\ref{algo:lda} both uses array $p_1(k)$. To reuse the data, as shown in Figure~\ref{figure:sample}, each sampler keeps a private index tree for $p_2(k)$. 

NVIDIA GPUs are equipped with L1 data cache and developers can decide which memory access instructions can access the cache. To further improve the performance, follow the performance models shown in ~\cite{iccad13xie}, we let the sparse matrix index access instructions to use the L1 cache to improve performance.

\subsubsection{Data compression}\label{sec:compress}. 
In addition to L1 data cache and shared memory, we also try to compress the data structure to improve the performance. The first optimization is precision compression. The corpus chunk and model $\theta_{D\times K}$ are stored in CSR format(a sparse matrix format). As the topic $K$ is smaller than $2^{16}$, we use the short integer as column index in the CSR format\footnote{\url{https://en.wikipedia.org/wiki/Sparse_matrix}}. Model $\phi_{K\times V}$ is a dense matrix, we also use short integer which is accurate enough. Another optimization is sub-expression reuse. We represent Eq.~\ref{eq:pk} in another format,

\begin{equation}\label{eq:pk}
\begin{split}
p^*(k) &= \frac{(\phi_{k,v} + \beta)}{\sum_{v\in[0,V)} \phi_{k,v} + \beta V}\\
p_1(k) &= \theta_{d,k}p^*(k)\ \ \ \ \  p_2(k) = \alpha p^*(k) 
\end{split}
\end{equation}

We find that $p_1(k)$ and $p_2(k)$ share the same sub-expression, $p^*(k)$. To reuse the common sub-expression, before start the sampling, each thread block first computes $p^*(k)$ and store it in the shared memory. Subsequent reads to $p^*(k)$ are all served by the shared memory, instead of GPU's off-chip memory.

\subsection{Model Update Optimization}\label{sec:update}

As shown in Figure~\ref{figure:overview}, after sampling each chunk, CuLDA\_CGS
updates the replicas of $\theta$ and $\phi$. Model $\phi$ is a dense matrix, the update algorithm is intuitive. We use the intrinsic atomic add instructions to update all elements of $\phi$. The corpus chunk is sorted in a word-first order, therefore, the update is word by word. Atomic function is expensive on processors. However, we observe that on NVIDIA GPUs, atomic functions that have good data locality shows good performance. Therefore, the update of $\phi$ is very fast. 
 
Updating model $\theta$ is more complex. $\theta$ is a sparse matrix and stored in CSR format, we can't directly use the atomic function. We design efficient update algorithm for model $\theta$. The update algorithm works document by document. In the first step, we first generate a dense array for the document, which corresponds to a line in $\theta$. We use the atomic functions in this step. In the second step, we transform the dense array to a sparse array in CSR format. In the first step, as the corpus chunk is in word-first order, we generate a document-word map to index all tokens in the same document. The map is generated on CPU's side at the data preprocessing stage. By doing so, we can enjoy the fast atomic functions. The second step uses the prefix sum array, which is a common algorithm for parallel processors. Please refer the source codes for the implementation details.

As shown in Section~\ref{sec:multi}, after the model update, we need the synchronize replicas of $\phi$ on different GPUs. Therefore, we first update model $\phi$. Hence, we can start the synchronization of model $\phi$ earlier and the update of model $\theta$ can be overlapped with the synchronization of model $\phi$.

\section{Experiment Results}\label{sec:exp}

\begin{table}[h]
    \begin{center}
      \caption{Configuration of the Evaluated Platforms.}\vspace{-0.5cm}
    \begin{tabular}{| >{\centering\arraybackslash}m{1.4 cm} | >{\centering\arraybackslash}m{6.2 cm} |}
    \hline
     \multicolumn{2}{|c|}{ \textbf{Maxwell Platform}} \\ \hline
     CPU & \vspace{0.06cm} Intel Xeon CPU E5-2670*2, 64 GB memory. \\[0ex] \hline
     GPU & \vspace{0.06cm} NVIDIA TITAN X GPU*1, Maxwell architecture, 336 GB/s.\\[0ex] \hline
	\multicolumn{2}{|c|}{\textbf{Pascal Platform}} \\ \hline
	 CPU & \vspace{0.06cm} Intel E5-2650 v3 processor*2, 64GB memory.\\[0ex] \hline
     GPU & \vspace{0.06cm} NVIDIA Titan Xp GPU*4, Pascal architecture, 550 GB/s memory. \\[0ex] \hline
      \multicolumn{2}{|c|}{\textbf{Volta Platform}} \\ \hline
      CPU & \vspace{0.06cm} IntelCPU E5-2690 v4 processor*2, 64GB Memory.\\[0ex] \hline
     GPU & \vspace{0.06cm} NVIDIA V100 GPU*2, 900GB/s. \\[0ex] \hline
  \end{tabular}
  \label{table:platform}
  \end{center}

\end{table}

\begin{table}[h]

   \begin{center}
      \caption{Details of workload data sets.}\vspace{-0.5cm}
    \begin{tabular}{|c |c |c |c|}
    \hline
     \textbf{Dataset}  & \textit{\#Tokens($T$)} & \textit{\#Documetns($D$)} & \textit{\#Words($V$)}       \\ \hline
      \textit{NYTimes}          & 99,542,125 &     299,752 & 101,636         \\ \hline
      \textit{PubMed}         & 737,869,083           & 8,200,000   & 141,043                  \\ \hline
  \end{tabular}
  \label{table:dataset}
  \end{center}

\end{table}

\begin{figure*}[h]
\centering
\includegraphics[scale=0.34,angle=0]{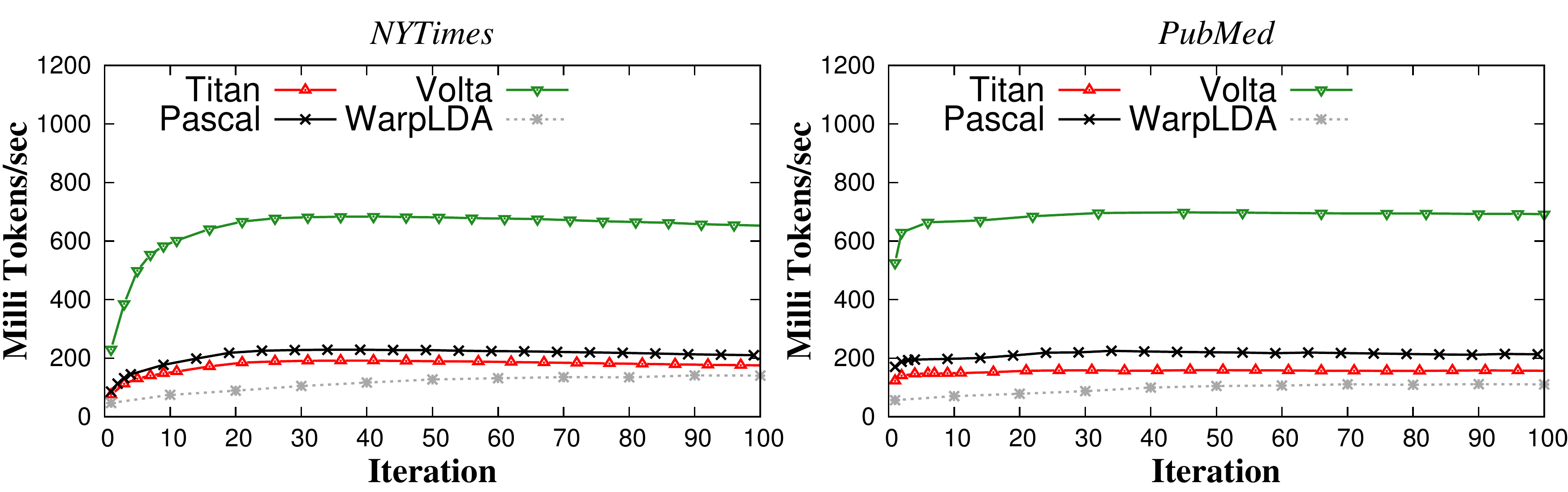}\vspace{-0.3cm}
\caption{Achieved sampling speed(\textit{\#Tokens/sec}) of \textit{CuLDA\_CGS} on different platforms(Titan, Pascal, Volta).}\label{figure:token}
\end{figure*}

\begin{figure*}[h]
\centering
\includegraphics[scale=0.34,angle=0]{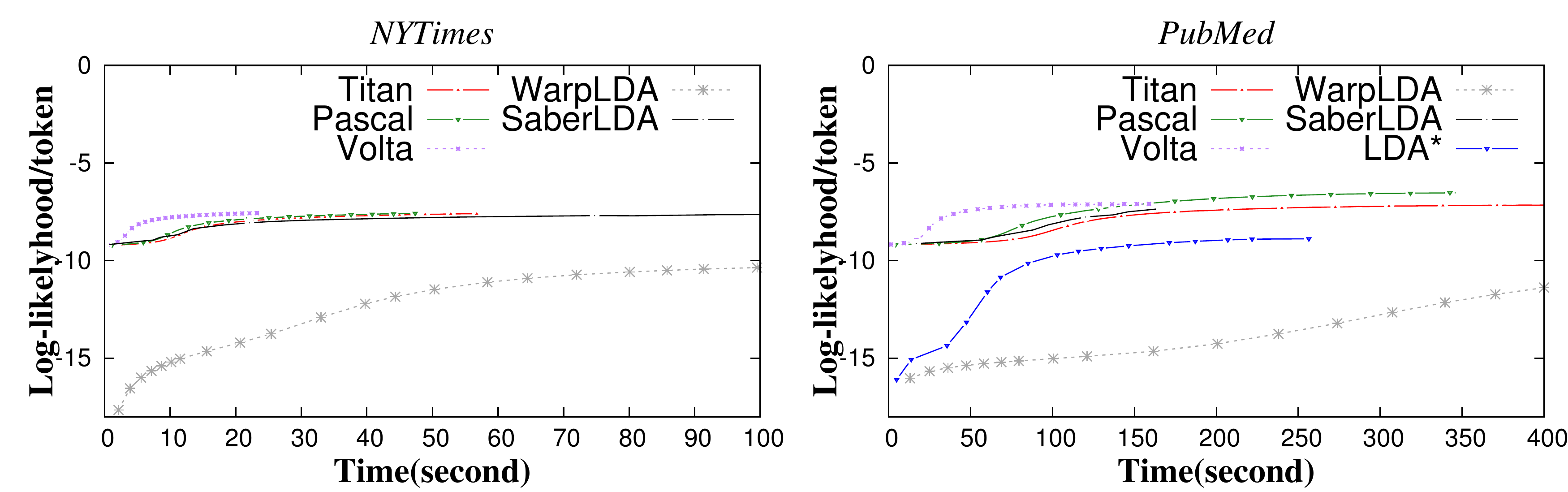}\vspace{-0.3cm}
\caption{Log-likelyhood per token w.r.t. time of evaluated LDA solutions.}\label{fig:loglike}
\end{figure*}

In this Section, we evaluate CuLDA\_CGS on real GPU platforms using wide-adopted data sets and compare it with the state-of-the-art LDA solutions. We use three commercially available GPUs as experimental platforms, the detailed configurations of the platforms are shown in Table~\ref{table:platform}. The GPUs and CPUs are connected via PCIe 3.0(up to 16GB/s). More clearly, the Volta Platform has the newest V100 GPUs and ideally, it delivers the best performance over other two platforms. We use two real data sets, \textit{NYTimes} and \textit{PubMed}, in the experiments. Those data sets are also used in previous works~\cite{vldb16warp,vldb17yu,kdd13cgs}. Table~\ref{table:dataset} shows the details of the two data sets. In this paper,  we use the same values of $\alpha$(50/$k$) and $\beta$(0.01) as previous works~\cite{asplos17saber}.


In the rest of this Section, we present the following experiments:
\begin{itemize}
    \item Performance. We evaluate CuLDA\_CGS on all three platforms and demonstrate that CULDA\_CGS is efficient and easy to scale to future GPU architectures.(Section~\ref{sec:performance})
    \item Comparison. We compare CuLDA\_CGS with the state-of-the-art LDA solutions, including single-node CPU-based solution, previous GPU-based soltion, and distributed system-based solution.(Section ~\ref{sec:compare})
    \item Multi-GPU scalability. We evaluate CuLDA\_CGS on multi-GPU systems and demonstrate that CuLDA\_CGS is able to scale to multiple GPUs for better performance.(Section ~\ref{sec:scale})
\end{itemize}

\subsection{Performance Evaluation}\label{sec:performance}

In this Section, we run CuLDA\_CGS on all three platforms and collect the $\#Tokens/sec$ metric for each iteration. Figure~\ref{figure:token} shows the performance results. We summarize the average $\#Tokens/sec$ of the first 100 iterations in Table~\ref{table:compare}.

\begin{table}[h]
   \begin{center}
      \caption{Average $\#Tokens/sec$ of CuLDA\_CGS and WarpLDA.}\label{table:compare}\vspace{-0.4cm}
    \begin{tabular}{|c |c |c |c|c|}
    \hline
     \textbf{Dataset}  & \textit{Titan} & \textit{Pascal} & \textit{Volta}    &\textit{WarpLDA}   \\ \hline
      \textit{NYTimes} & 173.6M  & 208.0M & 633.0M & 108.0M \\ \hline
      \textit{PubMed}  & 155.6M  & 213.0M & 686.2M & 93.5M \\ \hline
  \end{tabular}
  
  \end{center}
\end{table}

From Figure~\ref{figure:token}, the first observation is, the performance increases slowly at first few iterations and goes steady later. Previous work~\cite{vldb16warp} reports the similar observation on CPU platforms. The behind reason is, the sparsity rate of model $\theta_{D\times K}$ increases at the first few iterations. We also observe that, the performance variable of data set \textit{PubMed} is smaller than \textit{NYTimes}. That is because \textit{PubMed} has shorter average document length than \textit{NYTimes}(92 vs. 332). Therefore, the initial model sparsity rate of \textit{PubMed} is higher than that of \textit{NYTimes}. Hence, the initial performance of \textit{PubMed} is very close the the maximal performance.

Another observation is that CuLDA\_CGS is able to achieve the optimal performance on all three generations of GPUs. On average, compared with the Titan platform, CuLDA\_CGS achieves 1.28X speedup on the Pascal platform and 4.03X speedup on the Volta platform. The Titan GPU provides 336GB/s peak off-chip memory bandwidth and 24 processors on each GPU processors. In comparison, the Pascal GPU provides 550GB/s peak off-chip memory bandwidth and 28 processors. Therefore, CuLDA\_CGS can enjoy the increasing of hardware capability and achieves higher sampling speed. The Volta GPU is equipped with an even higher 900GB/s bandwidth and 80 processors. Therefore, the Volta GPU achieves the highest sampling speed. In conclusion, CuLDA\_CGS is able to scale to different GPU architectures and we believe it can be scaled to future GPUs as well.

Table~\ref{table:break} shows the performance breakdown of CuLDA\_CGS using data set NYTimes. We observe most of the execution time(79.4\%-87.9\%) is spent on the LDA sampling. It demonstrates that our model update algorithms shown in Section~\ref{sec:update} is efficient.

\begin{table}[h]
   \begin{center}
      \caption{Execution time breakdown of CuLDA\_CGS on data set \textit{NYTimes}.}\label{table:break}\vspace{-0.4cm}
    \begin{tabular}{|c |c |c |c|}
    \hline
     \textbf{Function}       & \textit{Titan} & \textit{Pascal} & \textit{Volta}    \\ \hline
     \textit{Sampling}      & 87.7\%  & 87.9\% & 79.4\% \\ \hline
     \textit{Update $\theta$} & 8.0\% & 9.3\% & 10.8\% \\ \hline 
     \textit{Update $\phi$}   & 4.3\%  & 1.7\%\% & 9.8\% \\ \hline
  \end{tabular}
  
  \end{center}
\end{table}

\subsection{Performance Comparison}\label{sec:compare}

To demonstrate CuLDA\_CGS is efficient, we compare CuLDA with the state-of-the-art LDA solutions. We use the \textit{log-likelyhood per token} as the performance metric. Figure~\ref{fig:loglike} shows the log-likehood per token of the evaluated solutions. The following techniques are evaluated:
\begin{itemize}
    \item CPU-based LDA solution~\cite{vldb16warp,www15light,kdd14alias,kdd09sparse}. We select the \textbf{WarpLDA} as it's the most updated solution. The source code is public available and we evaluate it on our Volta Platform. The used CPU is the most powerful one among all of the in-hand CPUs.
    \item Distributed LDA solution~\cite{www15nomad,yahoo12,vldb17yu,kdd15pet}. W select the most updated \textbf{LDA$^*$}. It employs up to 5600 machines to accelerate large-scale LDA and uses parameter server to manage the parallel workers. As the code is not available, we report their results on data set PubMed directly. For this data set, LDA$^*$ uses 20 nodes. 
    \item GPU-based LDA solution~\cite{canny2013bidmach,asplos17saber}. We select the most updated \textbf{SaberLDA}~\cite{asplos17saber} as comparison. SaberLDA is designed for single GPU and as the code is not publicly available, we cite the best reported performance in the paper. The cited results are conducted on NVIDIA GTX 1080 GPU that is at the same generation with our Titan platform and it's more powerful than Titan platform.
\end{itemize}

Compared with CPU-based solution WarpLDA~\cite{vldb16warp}, CuLDA\_CGS shows significant performance advantage. From Figure~\ref{figure:token} and Table~\ref{table:compare}, we observe that CuLDA\_CGS achieves much higher sampling speed(1.61X-7.34X) than WarpLDA on all used platforms and data sets. According to the original paper, one main performance advantage of WarpLDA is that it optimizes the cache to improve the memory efficiency. In comparison, CuLDA\_CGS relies on the software managed cache to achieve better performance. At the mean time, CuLDA\_CGS tries the reuse the sub-expression and compress data size to minimize memory footprint. In conclusion, CuLDA\_CGS is more efficient than WarpLDA.
 
Compared with distributed LDA solution, CuLDA\_CGS is more efficient. The main reason is, the distributed LDA solution is limited by the network bandwidth. As reported in the original paper~\cite{vldb17yu}, the machines used by LDA$^*$ are connected by 10Gb/s ethernet. Such a bandwidth is much slower than the PCIe bandwidth. As the LDA samplers need to synchronize the model after each iteration, the performance is limited by the network bandwidth. Note that CuLDA\_CGS only uses one GPU in this evaluation. It's more powerful when scaling to multiple GPUs. 


CuLDA\_CGS achieves faster convergence speed than existing GPU-based LDA solution, SaberLDA~\cite{asplos17saber}. First, CuLDA\_CGS achieves faster sampling speed. As reported in the paper, SaberLDA achieves $~$120M tokens/sec for NYTimes data set on a GTX 1080 GPU. In comparison, CuLDA\_CGS achieves 173.6M tokens/sec on a Titan X GPU. Note that GTX 1080 is more powerful as it is equipped with more hardware resources than Titan X GPU. Thus, CuLDA\_CGS is better than SaberLDA in terms of sampling speed with a lower-end GPU. Second, we have demonstrated that CuLDA\_CGS achieves high performance on different generations of GPUs while the cross platform results of SaberLDA is not reported. Third, in Section~\ref{sec:scale}, we will demonstrate that CuLDA\_CGS can benefit from multiple GPUs while SaberLDA lacks of multi-GPU support.

\subsection{Multi-GPU Scaling}\label{sec:scale}

\begin{figure}[h]
\centering
\includegraphics[scale=0.24,angle=0]{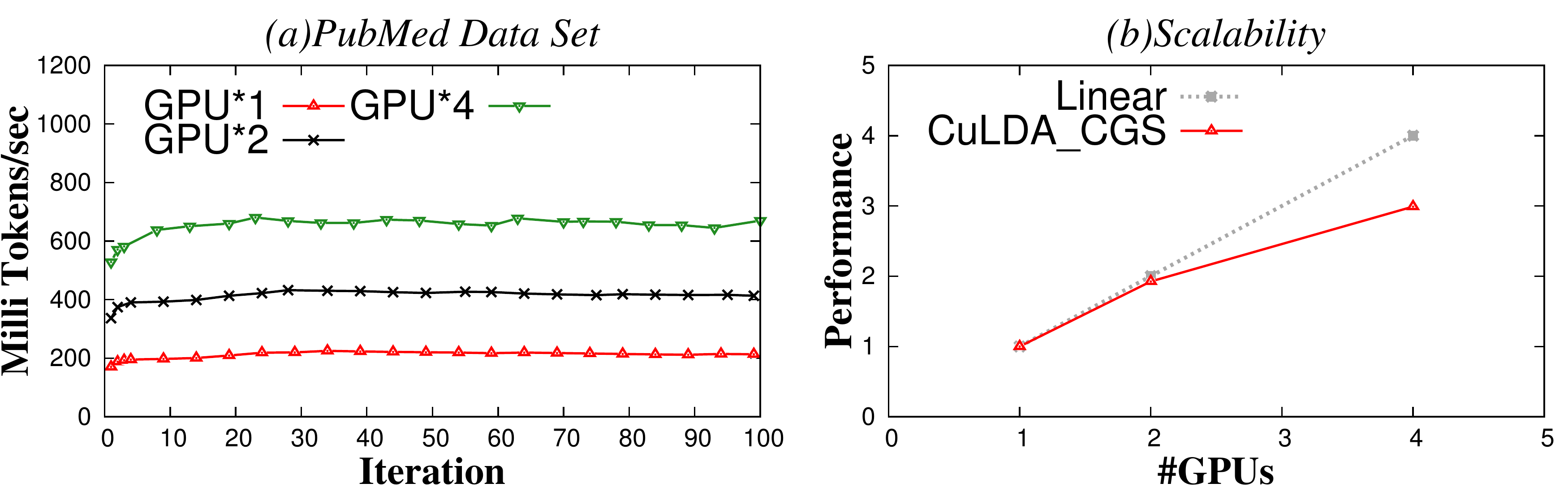}\vspace{-0.3cm}
\caption{Performance scalability evaluation of CuLDA\_CGS using PubMed data set on the Pascal platform.}\label{figure:scale}
\end{figure}

In this Section, we evaluate CuLDA\_CGS scalability on multiple GPUs. We evaluate CuLDA\_CGS on the Pascal platform using PubMed data set. Figure~\ref{figure:scale}(a) shows the detailed performance results and Figure~\ref{figure:scale}(b) shows the normalized performance of first 100 iterations. Compared with one GPU, CuLDA\_CGS achieves 1.93X and 2.99X speedup when using two and four GPUs, respectively. 

\section{Conclustion}\label{sec:con}

LDA is a popular model for large-scale document analysis. However, the computational complexity and large input data size make LDA solution time-consuming. In this paper, we present CuLDA\_CGS, an efficient and scalable solution to large-scale LDA problems. Our proposed CuLDA\_CGS is based on heterogeneous system that is composed CPUs and GPUs. On one hand, we design efficient workload partition and model synchronization algorithms to utilize the computational horsepower of multiple GPUs. On other hand, we design efficient LDA sampling algorithm to maximize the memory efficiency and minimize the model update overhead. We evaluate CuLDA\_CGS on three platforms using widely-adopted data sets. Evaluations show that CuLDA\_CGS is efficient and scalable to multiple GPUs. Compared with existing LDA solutions, CuLDA\_CGS shows significant performance advantage.


{\scriptsize
\bibliographystyle{abbrv}
\vspace{10pt}
\bibliography{main}
}

\end{document}